%% file: conference_101719.tex
\documentclass[conference]{IEEEtran}
\IEEEoverridecommandlockouts
\usepackage{cite}
\usepackage{amsmath,amssymb,amsfonts}
\usepackage{bm}
\usepackage{graphicx}
\usepackage{textcomp}
\usepackage{xcolor}
\usepackage{amsmath}
\usepackage[T1]{fontenc}
\usepackage{graphicx}
\usepackage{cite}
\usepackage{algorithm}
\usepackage{algpseudocode}
\usepackage{amsmath,amssymb,amsfonts}
\usepackage{amsthm}
\usepackage{textcomp}
\usepackage{xcolor}
\usepackage{subcaption}
\usepackage{upgreek}
\usepackage{booktabs}
\usepackage{balance}
\usepackage[hidelinks]{hyperref}
\def\BibTeX{{\rm B\kern-.05em{\sc i\kern-.025em b}\kern-.08em
    T\kern-.1667em\lower.7ex\hbox{E}\kern-.125emX}}

\makeatletter
\newcommand{\linebreakand}{%
  \end{@IEEEauthorhalign}
  \hfill\mbox{}\par
  \mbox{}\hfill\begin{@IEEEauthorhalign}
}

\newtheorem{thm}{Theorem}[section]

\newtheorem{rem}{Remark}

\begin{document}

\title{An $\mathcal{O}(\log_2N)$ SMC$^2$ Algorithm on Distributed Memory with an Approx. Optimal L-Kernel\\
\thanks{This research was funded in whole, or in part, by the Wellcome Trust [226691/Z/22/Z]. For the purpose of Open Access, the author has applied a CC BY public copyright license to any Author Accepted Manuscript version arising from this submission. CR was funded by the Wellcome CAMO-Net UK grant: 226691/Z/22/Z; JM was funded by a Research Studentship jointly funded by the EPSRC Centre for Doctoral Training in Distributed Algorithms EP/S023445/1; AV and SM were funded by EPSRC through the Big Hypotheses under Grant EP/R018537/1}
\thanks{\bf{\footnotesize \textcopyright 2023 IEEE. Personal use of this material is permitted.
  Permission from IEEE must be obtained for all other uses, in any current or future
  media, including reprinting/republishing this material for advertising or promotional
  purposes, creating new collective works, for resale or redistribution to servers or
  lists, or reuse of any copyrighted component of this work in other works.}}
}

\author{\IEEEauthorblockN{Conor Rosato\IEEEauthorrefmark{1}, Alessandro Varsi\IEEEauthorrefmark{2},
Joshua Murphy\IEEEauthorrefmark{2} and
Simon Maskell\IEEEauthorrefmark{2}}\\
\IEEEauthorblockA{\IEEEauthorrefmark{1} Department of Pharmacology and Therapeutics, University of Liverpool, United Kingdom \\
\IEEEauthorrefmark{2} Department of Electrical Engineering and Electronics, University of Liverpool, United Kingdom\\
Email: \{cmrosa, a.varsi, joshua.murphy, smaskell\}@liverpool.ac.uk}}


\maketitle

\input{0_Abstract}


\input{1_Introduction}

\input{3_PF}

\input{4_PMCMC}

\input{5_SMCsquared}

\input{6_Results}

\input{7_Conclusions}

\input{8_References}

\input{9_Appendix}

\end{document}

%% file: 0_Abstract.tex
\begin{abstract}

    Calibrating statistical models using Bayesian inference often requires both accurate and timely estimates of parameters of interest. Particle Markov Chain Monte Carlo (p-MCMC) and Sequential Monte Carlo Squared (SMC$^2$) are two methods that use an unbiased estimate of the log-likelihood obtained from a particle filter (PF) to evaluate the target distribution. P-MCMC constructs a single Markov chain which is sequential by nature so cannot be readily parallelized using Distributed Memory (DM) architectures. This is in contrast to SMC$^2$ which includes processes, such as importance sampling, that are described as \textit{embarrassingly parallel}. However, difficulties arise when attempting to parallelize resampling. None-the-less, the choice of backward kernel, recycling scheme and compatibility with DM architectures makes SMC$^2$ an attractive option when compared with p-MCMC. In this paper, we present an SMC$^2$ framework that includes the following features: an optimal (in terms of time complexity) $\mathcal{O}(\log_2N)$ parallelization for DM architectures, an approximately optimal (in terms of accuracy) backward kernel, and an efficient recycling scheme. On a cluster of $128$ DM processors, the results on a biomedical application show that SMC$^2$ achieves up to a $70\times$ speed-up vs its sequential implementation. It is also more accurate and roughly $54\times$ faster than p-MCMC. A GitHub link is given which provides access to the code.

\end{abstract}

%% file: 1_Introduction.tex
\section{Introduction}\label{sec:introduction}

In Bayesian inference problems, the posterior $p(\mathbf{\theta}|\mathbf{y})$ is only known up to a constant of proportionality such that the target, $\pi(\bm{\theta}) = p(\mathbf{\theta})p(\mathbf{y}|\mathbf{\theta}) \propto p(\mathbf{\theta}|\mathbf{y})$,
where $\mathbf{y}$ is the data, $p(\mathbf{y}|\bm{\theta})$ is the likelihood and $p(\bm{\theta})$ is the prior. Two methods for sampling from a static $\pi(\bm{\theta})$ include: Markov Chain Monte Carlo (MCMC) and Sequential Monte Carlo (SMC) samplers \cite{del2006sequential}. Theoretically, sampling with a single Markov chain using MCMC is ergodic and ensures convergence to the correct posterior distribution. However, sampling with one MCMC chain cannot readily exploit parallel computing architectures such as Message Passing Interface (MPI). This is due to the next MCMC sample being dependent on the current sample. Running multiple MCMC chains in parallel is a method for overcoming this issue \cite{altekar2004parallel}. The challenges of parallelizing a single MCMC chain contrasts with SMC samplers which have inherent steps that can be described as \textit{embarrassingly parallel}. Sampling $N$ independent samples from a proposal distribution is one such example. An Importance Sampling (IS) step weights these particles, with an L-kernel minimizing the variance of the weights associated to the samples. A recent advancement outlined in \cite{green2022increasing} defines an approximately optimal L-kernel which reduces the variance by 99\% in certain scenarios. 

The standard proposal used in MCMC and SMC samplers is Random Walk (RW) \cite{hastings1970monte} and \cite{robert2010metropolis}. If gradients can be calculated, gradient based MCMC proposals such as Hamiltonian Monte Carlo (HMC) \cite{neal2012mcmc} and the No-U-Turn Sampler (NUTS) \cite{NUTS} can be employed. Probabilistic programming languages (ppls) such as PyMC3 \cite{pymc3} and Stan \cite{carpenter2017stan} use NUTS as the default sampler. 

A common approach when estimating dynamic posteriors is to use a Particle Filter (PF) \cite{arulampalam2002tutorial}. Similar to an SMC sampler, a PF draws a set of particles from a proposal distribution but performs IS sequentially in a process termed Sequential Importance Sampling (SIS). In both PF and SMC samplers, a resampling step is included to avoid particle degeneracy \cite{arulampalam2002tutorial}. It does this by replicating and eradicating particles with higher and lower importance weights, respectively. Multiple resampling algorithms exist with the most popular outlined in \cite{resamplingmethods}. IS, which is \textit{embarrassingly parallel}, has a time complexity of $\mathcal{O}(1)$ for $P = N$ processing elements (or processors, the terms are used interchangeably). However, resampling incurs a computational cost of $\mathcal{O}(N)$ and has been thought of as a bottleneck when employing parallelization due to message passing.

Examples of parallel resampling using shared memory on GPU and CPU architectures can be found here \cite{Lopez, Lopez2, Murray_Multiresampling, Alessandro2}. To make use of High Performance Computing (HPC) applications, Distributed Memory (DM) architectures need to be employed. When compared to shared memory, DM can be more challenging to parallelize. This is due to cores not having access to other cores memory without exchanging messages. A recent advancement in parallel resampling using DM, that improves on the work in \cite{Alessandro} and \cite{Alessandro3}, can be found in \cite{Alessandro5} which achieves asymptotically optimal time complexity $\mathcal{O}(\log_2N)$.

As the number of particles in a PF tends to infinity, it is proven that the summation of the log-weights at each time-step is an unbiased estimate of the log-likelihood \cite{andrieu_doucet_holenstein_2010}. Two methods for parameter estimation that utilize this estimate of the log-likelihood is Particle-Markov Chain Monte Carlo (p-MCMC) \cite{andrieu_doucet_holenstein_2010} and SMC Squared (SMC$^2$) \cite{chopin2013smc2}. P-MCMC can be prohibitive to use in practice if a single run of the PF is computationally expensive as a log-likelihood evaluation will need to be made at each MCMC iteration. Similarly to MCMC, parallelism is not readily available. SMC$^2$ has two layers: an SMC sampler with a set of samples used for parameter estimation with each particle evaluating the log-likelihood using a PF. Like the SMC sampler, SMC$^2$ can be  parallelized. We note that SMC$^2$ has been parallelized when modelling specific applications \cite{carson2018bayesian, ABM, lux2023sequential}. However, a general purpose fully parallelized SMC$^2$ framework is yet to exist.

Recycling allows for the approximations of $\pi(\bm{\theta})$ expectations from all iterations of the SMC sampler instead of an estimate from the last iteration. Recycling has been seen to improve estimates in \cite{green2022increasing, nguyen2015efficient}. To the authors' knowledge, recycling has yet to be used within SMC$^2$.

In this paper, we present an SMC$^2$ framework that has an optimal $\mathcal{O}(\log_2N)$ time complexity parallelization for DM, an approximately optimal L-kernel and an efficient recycling scheme, all used in the context of SMC$^2$ for the first time. We demonstrate it to be faster and more accurate than p-MCMC. The structure of this paper is as follows: The PF is described in Section~\ref{sec:particle_filter}. P-MCMC is briefly outlined in Sections~\ref{sec:Pmcmc}. The proposed SMC$^2$ along with its parallelization is described in Section \ref{sec:SMC_sampler} and Appendix \ref{app:parallel_resamp}, respectively. Section~\ref{sec:Examples} provides the numerical results and concluding remarks are presented in Section~\ref{sec:conclusions}.

%% file: 3_PF.tex
\section{Particle Filter}\label{sec:particle_filter}

State-Space Models (SSMs) have been used to model dynamic systems in a wide range of research fields \cite{doucet2001sequential}. SSMs are a favored method for representing the dependence of latent states in non-linear dynamical systems. A SSM consists of a state equation, 
\begin{equation}\label{xt}
\mathbf{x}_{t} \mid \mathbf{x}_{t-1} \sim f(\mathbf{x}_{t} \mid \mathbf{x}_{t-1}, \bm{\theta}),
\end{equation}
\noindent which is parameterized by $\bm{\theta}$ and models how the dynamical system moves from the previous state to the current state, $\mathbf{x}_t$, at time $t$. They also include an observation equation 
\begin{equation}\label{yt}
\mathbf{y}_{t} \mid \mathbf{x}_{t} \sim f(\mathbf{y}_{t} \mid \mathbf{x}_{t}, \bm{\theta}),
\end{equation}
\noindent which describes how the observation, $\mathbf{y}_t$, is linked to the state equation.

The PF uses a set of $N_x$ particles to recursively represent any nonlinear, non-Gaussian SSM as $N_x\to\infty$. At every time step $t$, particles are drawn from a proposal distribution, $q\left(\mathbf{x}_{1:t}|\mathbf{y}_{1:t},\bm{\theta}\right)$, which is parameterized by the sequence of states $\mathbf{x}_{1:t}$ and measurements $\mathbf{y}_{1:t}$. Different options exist when constructing the proposal \cite{rosato2022efficient}. The $i'$-th particle, $\mathbf{x}_{1:t}^{i'}$, $\forall i' \in \{1...N_x\}$, has an associated weight, $\mathbf{w}_{t}^{i'}$, which indicates its contribution to any estimates\footnote{We use the notation $i'$ to iterate over the particles of the PF and to differentiate between samples in the SMC sampler.}. The proposal of the PF can be constructed as $q\left(\mathbf{x}_1|\mathbf{y}_1,\bm{\theta}\right)\prod_{\tau=2}^t q\left(\mathbf{x}_\tau|\mathbf{x}_{\tau-1},\mathbf{y}_\tau,\bm{\theta}\right)$ such that an estimate with respect to the joint distribution, $p\left(\mathbf{y}_{1:t},\mathbf{x}_{1:t}|\bm{\theta}\right)$, can be posed as follows, with $f\left(\mathbf{x}_{1:t}^{i'}\right)$ denoting the function of interest on $\mathbf{x}_{1:t}^{i'}$:
\begin{equation}
\int p\left(\mathbf{y}_{1:t},\mathbf{x}_{1:t}|\bm{\theta}\right)f\left(\mathbf{x}_{1:t}\right) d\mathbf{x}_{1:t} \approx \frac{1}{N_x}\sum\nolimits_{i'=1}^{N_x} \mathbf{w}_{t}^{i'}f\left(\mathbf{x}_{1:t}^{i'}\right).\label{eq:jointexp}
\end{equation}
This is an unbiased estimate, where (for $t>$1)
\begin{align}
\mathbf{w}_{t}^{i'}=&\frac{p\left(\mathbf{y}_1|\mathbf{x}_1^{i'},\bm{\theta}\right)p\left(\mathbf{x}_1^{i'}|\bm{\theta}\right)}{q\left(x^{i'}_1|\mathbf{y}_1,\bm{\theta}\right)}\nonumber\\
&\times \frac{\prod_{\tau=2}^{t}p\left(\mathbf{y}_\tau|\mathbf{x}_\tau^{i'},\bm{\theta}\right)p\left(\mathbf{x}_\tau^{i'}|x^{i'}_{\tau-1},\bm{\theta}\right)}{\prod_{\tau=2}^t q\left(\mathbf{x}_\tau^{i'}|\mathbf{x}_{\tau-1}^{i'},\mathbf{y}_\tau,\bm{\theta}\right)} \\ 
=& \mathbf{w}_{t-1}^{i'}\frac{p\left(\mathbf{y}_t|\mathbf{x}_t^{i'},\bm{\theta}\right)p\left(\mathbf{x}_t^{i'}|x^{i'}_{t-1},\bm{\theta}\right)}{q\left(\mathbf{x}_t^{i'}|\mathbf{x}_{t-1}^{i'},\mathbf{y}_t\right)},\label{eq:weightupdate}
\end{align}
and is a recursive formulation for the unnormalized weight, $\mathbf{w}_{1:t}^{i'}$. A (biased) estimate with respect to the posterior, $p\left(\mathbf{x}_{1:t}|\mathbf{y}_{1:t},\bm{\theta}\right)$ can be made by 
\begin{align}
\int p\left(\mathbf{x}_{1:t}|\mathbf{y}_{1:t},\bm{\theta}\right)&f\left(\mathbf{x}_{1:t}\right) d\mathbf{x}_{1:t} \\
= &\int \frac{p\left(\mathbf{y}_{1:t},\mathbf{x}_{1:t}|\bm{\theta}\right)}{p\left(\mathbf{y}_{1:t}|\bm{\theta}\right)}f\left(\mathbf{x}_{1:t}\right) d\mathbf{x}_{1:t},
\end{align}
where 
\begin{equation}
p\left(\mathbf{y}_{1:t}|\bm{\theta}\right) = \int  p\left(\mathbf{y}_{1:t},\mathbf{x}_{1:t}|\bm{\theta}\right)d\mathbf{x}_{1:t}\approx  \frac{1}{N_x}\sum\nolimits_{i'=1}^{N_x} \mathbf{w}_{t}^{i'}.\label{eq:likelihood_pf}
\end{equation}
This is in line with the joint distribution in \eqref{eq:jointexp} such that 
\begin{align}
\int  p&\left(\mathbf{x}_{1:t}|\mathbf{y}_{1:t},\bm{\theta}\right)f\left(\mathbf{x}_{1:t}\right)d\mathbf{x}_{1:t} \nonumber \\
& \approx \frac{1}{\frac{1}{N_x}\sum\nolimits_{i'=1}^{N_x} {\mathbf{w}}_{t}^{i'}}
\frac{1}{N_x}\sum\nolimits_{i'=1}^{N_x} {\mathbf{w}}_{t}^{i'}f\left(\mathbf{x}_{1:t}^{i'}\right)  \\ 
&= \sum\nolimits_{i'=1}^{N_x} \tilde{\mathbf{w}}_{t}^{i'}f\left(\mathbf{x}_{1:t}^{i'}\right),
\label{eq:a}
\end{align}
where
\begin{equation}
\tilde{\mathbf{w}}_{t}^{i'} = \frac{\mathbf{w}_{t}^{i'}}{\sum\nolimits_{i'=1}^{N_x} \mathbf{w}_{t}^{i'}}, \ \ \forall i.
\label{eq:normalisedweights}
\end{equation}
are the normalized weights. 

As time evolves, \eqref{eq:normalisedweights} becomes increasingly skewed such that one of the weights becomes close to unity, while the others approach zero. This is known as particle degeneracy and can be mitigated by resampling. The goal of resampling is to delete particles that have low weights and replace them with copies of particles with high weights. The probability of selecting each resampled particle is proportional to its normalized weight. Monitoring the number of effective particles, ${N_x}_{eff}$, given by 
\begin{align} \label{eq: ess}
{N_x}_{eff} = \frac{1}{\sum\nolimits_{i'=1}^{{N_x}} \left(\tilde{\mathbf{w}}_{t}^{i'}\right)^2},
\end{align}
and resampling when ${{N_x}}_{eff}$ is less than ${{N_x}}/2$ is a common approach for determining when to resample. Several resampling schemes have been proposed in the literature~\cite{resamplingmethods}. Multinomial resampling is chosen within the PF. Multinomial resampling involves drawing from the current particle set ${N_x}$ times, proportionally to the associated weights of the particles. The associated distribution is defined by
$\tilde{\mathbf{w}}_{t}^{i'}, \ \  \forall i$. To keep the total unnormalized weight constant (such that the approximation \eqref{eq:likelihood_pf} is the same immediately before and after resampling), each newly-resampled sample is assigned an unnormalized weight
\begin{equation}
\frac{1}{N_x}\sum\nolimits_{i'=1}^{N_x} \mathbf{w}_{t}^{i'},
\label{eq:normalisedafterresampling}
\end{equation}
such that the normalized weights after resampling are $\frac{1}{N_x}$.

%% file: 4_PMCMC.tex
\section{Particle-Markov Chain Monte Carlo}\label{sec:Pmcmc}
    
    P-MCMC combines MCMC and a PF to obtain numerical estimates related to $\pi(\bm{\theta})$ for which exact inference is intractable. The original contribution of \cite{andrieu_doucet_holenstein_2010} uses a PF to calculate an unbiased estimate of the marginal log-likelihood, $p(\mathbf{y}_{1:T}|\bm{\theta})$. The log-likelihood can be approximated by summing the unnormalized PF weights in \eqref{eq:likelihood_pf} for $t=1,..,T$. This is a byproduct of running the PF and so no additional calculations are required. The resulting log-likelihood estimate can be used within the Metropolis Hastings Random Walk (MHRW) algorithm and allows the acceptance probability to be formulated as
    \begin{equation}
        \alpha(\bm{\theta},\bm{\theta}') = \min \left\{1, \frac{p\left(\bm{\theta}^{\prime}\right)p(\mathbf{y}_{1:T}|\bm{\theta}')q(\bm{\theta} |  \bm{\theta}')}{p(\bm{\theta})p(\mathbf{y}_{1:T}|\bm{\theta})q(\bm{\theta}' | \bm{\theta})}\right\},
        \label{eq:MHalgorithm_}
    \end{equation}
    where $p(\bm{\theta})$ is the prior density and $q(\bm{\theta}' | \bm{\theta})$ the proposal. Reference \cite{andrieu_doucet_holenstein_2010} prove that the Markov chain converges to $\pi(\bm{\theta})$ when using a fixed number of ${N_x}$ particles. Estimates are computed as the mean of the samples.

%% file: 5_SMCsquared.tex
\section{SMC$^2$}\label{sec:SMC_sampler}
    At each iteration $k$, SMC$^2$ targets $\pi(\mathbf{\bm{\theta}})$, where $\mathbf{\bm{\theta}} \in \mathbb{R}^{D}$, via IS and resampling steps which are outlined in sections~\ref{sec:importance_sampling} and \ref{sec:smcsampler_resampling}, respectively. The joint distribution from all states until $k=K$ is defined to be
    \begin{equation}
        \pi(\mathbf{\bm{\theta}}_{1:K}) = \pi(\mathbf{\bm{\theta}}_{K}) \prod_{k=2}^{K} L(\mathbf{\bm{\theta}}_{k-1} | \mathbf{\bm{\theta}}_{k}),
    \end{equation}
    where $L(\mathbf{\bm{\theta}}_{k-1} | \mathbf{\bm{\theta}}_{k})$ is the L-kernel, which is a user-defined probability distribution. The choice of this distribution can affect the efficiency of the sampler \cite{green2022increasing}.

    \subsection{Importance Sampling}\label{sec:importance_sampling}
        At $k=1$, $N$ samples $\forall i = 1, \dots, N$ are drawn from a prior distribution $q_1(\cdot)$ as follows:
        \begin{equation} \label{eq: init_sample}
            \mathbf{\bm{\theta}}^i_1 \sim q_1(\cdot), \ \ \forall i,
        \end{equation}
        and weighted according to
        \begin{equation} 
            \mathbf{w}^i_1 = \frac{\pi(\mathbf{\bm{\theta}}^i_1)}{q_1(\mathbf{\bm{\theta}}^i_1)}, \ \ \forall i\label{init_weights}.
        \end{equation}
        As described for p-MCMC in Section~\ref{sec:Pmcmc}, $\pi(\mathbf{\bm{\theta}}^i_k)$ in \eqref{init_weights} is the prior $p(\bm{\theta}^{i}_{k})$ multiplied by the likelihood $p(\mathbf{y}_{1:T}|\bm{\theta}^{i}_{k})$ (which is given by the PF).
        
        At $k>1$, subsequent samples are proposed based on samples from the previous iteration via a proposal distribution, $q(\mathbf{\bm{\theta}}^i_k|\mathbf{\bm{\theta}}^i_{k-1})$, as follows: 
        \begin{equation} \label{eq: proposal}
            \mathbf{\bm{\theta}}^i_k \sim q(\cdot|\mathbf{\bm{\theta}}^i_{k-1}).
        \end{equation}
        The proposal is commonly chosen to be Gaussian with a mean of $\mathbf{\bm{\theta}}^i_{k-1}$ and a covariance of $\mathbf{\Sigma} \in \mathbb{R}^{DxD}$, such that
        \begin{equation}
            q(\mathbf{\bm{\theta}}^i_k|\mathbf{\bm{\theta}}^i_{k-1}) = \mathcal{N}(\mathbf{\bm{\theta}}^i_k; \mathbf{\bm{\theta}}^i_{k-1}, \Sigma), \ \ \forall i.
        \end{equation}
        These samples are weighted according to 
        \begin{equation}
            \mathbf{w}^i_{k} = \mathbf{w}^i_{k-1} \frac{\pi(\mathbf{\bm{\theta}}^i_{k})}{\pi(\mathbf{\bm{\theta}}^i_{k-1})} \frac{L(\mathbf{\bm{\theta}}^i_{k-1}|\mathbf{\bm{\theta}}^i_{k})}{q(\mathbf{\bm{\theta}}^i_{k}|\mathbf{\bm{\theta}}^i_{k-1})}, \ \ \forall i.\label{eq:l_weights}
        \end{equation}
        Estimates of the expectations of functions, such as moments, on the distribution are realised by 
        \begin{equation} 
            \Tilde{\mathbf{f}}_{k} = \sum\nolimits_{i=1}^{N} \Tilde{\mathbf{w}}^i_{k} \mathbf{\bm{\theta}}^i_{1:k}, \label{realised_estimates}
        \end{equation}
        where the normalised weights $\Tilde{\mathbf{w}}^i_{k}$ are calculated as in \eqref{eq:normalisedweights}:
        \begin{equation}
            \tilde{\mathbf{w}}_{k}^{i} = \frac{\mathbf{w}_{k}^{i}}{\sum\nolimits_{i=1}^{N} \mathbf{w}_{k}^{i}}, \ \ \forall i.
            \label{eq: normalise_smc2}
        \end{equation} 
        Recycling constants are also calculated (see Section \ref{sec:recycling}):
        \begin{equation}
                \mathbf{c}_k = \frac{\mathbf{l}_k}{\sum\nolimits_{k=1}^K \mathbf{l}_k}, 
                \ \ \forall k,\label{eq: opt_recycling_constant}
            \end{equation}
            where 
            \begin{align} \label{eq: recycling_I}
                \mathbf{l}_k = \frac{\left(\sum\nolimits_{i=1}^{{N}} 
                \mathbf{w}_{k}^{i}\right)^2}{\sum\nolimits_{i=1}^{{N}} 
                \left(\mathbf{w}_{k}^{i}\right)^2}.
                \ \ \forall k.
            \end{align}
        Sample degeneracy occurs for the same reasoning as the PF and is mitigated by resampling.
        \subsection{Resampling}\label{sec:smcsampler_resampling}
            SMC$^2$ computes the Effective Sample Size (ESS) in an analogous way to \eqref{eq: ess}: 
            \begin{align} \label{eq: ess_smc2}
                {N}_{eff} = \frac{1}{\sum\nolimits_{i=1}^{{N}} \left(\tilde{\mathbf{w}}_{k}^{i}\right)^2}.
            \end{align}
            Resampling is undertaken if $N_{eff} < N/2$. However, systematic is used in place of multinomial resampling due to the variance of the normalised weights after resampling being minimised. Systematic resampling is executed in three steps:
            
            \textit{Step 1 - Choice}. The cumulative density function, $\mathbf{cdf} \in \mathbb{R}^N$, of the normalised weights is computed as follows:
            \begin{equation} \label{eq: cdf}
                \mathbf{cdf}^i = N\sum\nolimits_{j=1}^{i} \Tilde{\mathbf{w}}_k^j, \ \ \forall i.
            \end{equation}
            A random variable, $u \sim U[0, 1]$, is drawn such that
            \begin{equation} \label{eq: ncopies}
                \mathbf{ncopies}^i = \lceil \mathbf{cdf}^{i+1} - u\rceil - \lceil \mathbf{cdf}^i - u\rceil, \ \ \forall i,
            \end{equation}
            represents the number of times the $i$-th sample has to be duplicated. Note samples for which $\mathbf{ncopies}^i = 0$ will be deleted. From \eqref{eq: ncopies}, it is straightforward to infer that:
            \begin{subequations} \label{eq: ncopies_definition}
                \begin{align}
                    0 \leq \mathbf{ncopies}^i &\leq N, \ \ \forall i, \label{eq: ncopies_range}\\
                    \sum\nolimits_{i=1}^{N} \mathbf{ncopies}^i &= N. \label{eq: workload}
                \end{align}
            \end{subequations}
            
            \textit{Step 2 - Redistribution}. This step involves copying and pasting the number of samples outlined in \eqref{eq: ncopies}. A textbook Sequential Redistribution (S-R) is described in Algorithm~\ref{alg: sequential_redistribution}. 
            \begin{algorithm}
                \small
                \caption{Sequential Redistribution (S-R)}\label{alg: sequential_redistribution}
                \textbf{Input: } $\mathbf{\bm{\theta}}$, $\mathbf{ncopies}$, $N$ \\
                \textbf{Output: } $\mathbf{\bm{\theta}}_{new}$
                \begin{algorithmic}[1]
                    \State $i \gets 1$
	            \For{$j\gets 1$; $j \leq N$; $j\gets j+1$}
                        \For{$copy\gets 1$; $copy \leq \mathbf{ncopies}^j$; $copy\gets copy+1$}
                            \State $\mathbf{\bm{\theta}}^{i}\gets \mathbf{\bm{\theta}}^j$
                            \State $i\gets i+1$
                        \EndFor
                    \EndFor
                \end{algorithmic}
            \end{algorithm}
            
            \textit{Step 3 - Reset}. After redistribution, all normalised weights are reset to $\frac{1}{N}$.
            This process occurs for $\forall k = 1, 2, \dots, K$ iterations. Once $k=K$, the estimates of $\mathbf{\bm{\theta}}_{1:K}$ are recycled according to the method outlined in Section \ref{sec:recycling}. Table~\ref{tab: smc2_tasks} outlines the time complexity of each process of SMC$^2$ on DM for $P$ processors. Appendix~\ref{app:parallel_resamp} provides a more in depth description of each process.
    \begin{table}[ht!] 
    \begin{center}
        \centering
        \small
        \caption{Time complexity of each process of SMC$^2$ on DM.\label{tab: smc2_tasks}}
        \begin{tabular}{cc}
            \toprule
            \toprule
            \textbf{\shortstack{Task name - Details}} & \textbf{\shortstack{Time complexity}}\\
            \midrule
            IS - Eq. \eqref{eq: proposal} \& \eqref{eq:l_weights}		& $\mathcal{O}(\frac{N}{P}) \rightarrow \mathcal{O}(1)$\\
            Normalise - Eq. \eqref{eq: normalise_smc2}			& $\mathcal{O}(\frac{N}{P} + \log_2P) \rightarrow \mathcal{O}(\log_2N)$\\
            ESS - Eq. \eqref{eq: ess_smc2}		& $\mathcal{O}(\frac{N}{P} + \log_2P) \rightarrow \mathcal{O}(\log_2N)$\\
            Choice - Eq. \eqref{eq: cdf} \& \eqref{eq: ncopies}			& $\mathcal{O}(\frac{N}{P} + \log_2P) \rightarrow \mathcal{O}(\log_2N)$\\
            Redistribution - Alg. \ref{alg: parallel_redistribution} & $\mathcal{O}(\frac{N}{P} + \frac{N}{P}\log_2P) \rightarrow \mathcal{O}(\log_2N)$\\
            Reset - $\mathbf{w}_k^i \gets \frac{1}{N} \forall i$		& $\mathcal{O}(\frac{N}{P}) \rightarrow \mathcal{O}(1)$\\
            Estimate - Eq. \eqref{realised_estimates}			& $\mathcal{O}(\frac{N}{P} + \log_2P) \rightarrow \mathcal{O}(\log_2N)$\\
            Recycling - Eq. \eqref{eq: recycling_I}			& $\mathcal{O}(\frac{N}{P} + \log_2P) \rightarrow \mathcal{O}(\log_2N)$\\
            \bottomrule
            \bottomrule
        \end{tabular}
    \end{center}
\end{table}
        \subsection{Recycling}\label{sec:recycling}
            In using \eqref{realised_estimates} to calculate estimates, only samples from the most recent iteration are used. Estimates from previous iterations can be utilized through
            \begin{equation}
                \Tilde{\mathbf{f}} = \sum\nolimits_{k=1}^K \mathbf{c}_k\Tilde{\mathbf{f}}_k,
            \end{equation}
            where $\sum\nolimits_{k=1}^K \mathbf{c}_k = 1$. The optimal recycling constants in \eqref{eq: opt_recycling_constant} maximize the ESS of the whole population of samples.

        \subsection{L-kernels}\label{sec:l_kernels}
            A common, suboptimal approach to selecting the L-kernel in \eqref{eq:l_weights} is to choose the same distribution as the forwards proposal distribution:
            \begin{equation}
                L(\mathbf{\bm{\theta}}^i_{k-1}|\bm{\theta}^i_{k}) = q(\mathbf{\bm{\theta}}^i_{k-1}|\mathbf{\bm{\theta}}^i_{k}), \ \ \forall i,
            \end{equation}
            such that, when $q(\cdot | \cdot)$ is symmetric, the weight update in \eqref{eq:l_weights} simplifies to
            \begin{equation}
                \mathbf{w}^i_{k} = \mathbf{w}^i_{k-1} \frac{\pi(\mathbf{\bm{\theta}}^i_{k})}{\pi(\mathbf{\bm{\theta}}^i_{k-1})}, \ \ \forall i.\label{eq:aimplified_l_weights}
            \end{equation}
            The asymptotically optimal L-kernel, outlined in \cite{del2006sequential}, minimizes the variance of the realized estimates in \eqref{realised_estimates}. The general form of this optimal L-kernel is often intractable, so an approximate, implementable form can be found in \cite{green2022increasing}. This approximate form involves estimating the joint distribution of $\bm{\theta}_{k-1}$ and $\bm{\theta}_{k}$ as a Gaussian:
            \begin{equation}
                \hat{q}(\mathbf{\bm{\theta}}_{k-1}, \mathbf{\bm{\theta}}_{k}) = \mathcal{N}\left(\begin{bmatrix}
                \mathbf{\bm{\theta}}_{k-1} \\
                \mathbf{\bm{\theta}}_{k}
                \end{bmatrix};
                \begin{bmatrix}
                    \mathbf{\bm{\mu}}_{k-1} \\
                    \mathbf{\bm{\mu}}_{k}
                \end{bmatrix},
                \begin{bmatrix}
                    \mathbf{\Sigma}_{k-1,k-1} & \mathbf{\Sigma}_{k-1,k} \\
                    \mathbf{\Sigma}_{k,k-1} & \mathbf{\Sigma}_{k,k}
                \end{bmatrix}\right),\label{Gauss_joint}
            \end{equation}
            which can then be taken as the marginal of this approximate joint distribution
            \begin{equation}
                L(\mathbf{\bm{\theta}}^i_{k-1}|\mathbf{\bm{\theta}}^i_{k}) = \hat{q}(\mathbf{\bm{\theta}}_{k-1}|\mathbf{\bm{\theta}}_{k}) = \mathcal{N}(\mathbf{\bm{\theta}}^i_k; \mathbf{\bm{\mu}}_{k-1|k}, \mathbf{\Sigma}_{k-1|k}), \ \ \forall i,\label{marginal_l}
            \end{equation}
            where 
            \begin{equation}
                \mathbf{\bm{\mu}}_{k-1|k}=\mathbf{\bm{\mu}}_{k-1}+\mathbf{\Sigma}_{k,k}^{-1} \mathbf{\Sigma}_{k,k-1}(\mathbf{\bm{\theta}}_k-\mathbf{\bm{\mu}}_k),
            \end{equation}
            and
            \begin{equation}
                \mathbf{\Sigma}_{k-1,k}=\mathbf{\Sigma}_{k-1,k-1}-\mathbf{\Sigma}_{k-1}\mathbf{\Sigma}_{k,k}^{-1}\mathbf{\Sigma}_{k,k-1}.
            \end{equation}

%% file: 6_Results.tex
\section{Numerical Results}\label{sec:Examples}
In this section, we compare SMC$^2$ and p-MCMC in terms of estimation accuracy and run-time for the same problem size (i.e., total number of samples). The model for comparison is a disease model for biomedical applications and is described in Section \ref{sec:sir}.

\subsection{Disease Model} \label{sec:sir}
The Susceptible, Infected and Recovered (SIR) \cite{kermack1927contribution}  epidemiological model simulates the spread of disease through a population of size $N_{pop}$. The population is split into three unobservable compartments. At each time step, $t$, proportions of the population move from the susceptible to infected and infected to recovered compartments. 

A method for incorporating stochasticity in the SIR model is to specify the number of individuals leaving a compartment as a binomial distributed random variable. The binomial probability, $p$, is defined as the rate at which individuals leave a compartment, given by
\begin{align}
p(\mathbf{S}\rightarrow \mathbf{I})&=1-e^{\frac{-\beta \mathbf{I}_{t-1} \mathbf{S}_{t-1}}{N_{pop}}}, \label{discrete_s_p}\\
p(\mathbf{I}\rightarrow \mathbf{R})&=1-e^{-\gamma}\label{discrete_i_p},
\end{align}
where $p(\mathbf{S}\rightarrow \mathbf{I})$ and $p(\mathbf{I}\rightarrow \mathbf{R})$ denote the probability of leaving the susceptible and infected compartments, respectively. The effective transmission rate of the disease and the mean recovery rate are given by $\beta$ and $\gamma$, respectively. 
The corresponding binomial distributions are
\begin{align}
n(\mathbf{S}\rightarrow \mathbf{I})&\sim \text{Binomial}(\mathbf{S}_{t-1}, p(\mathbf{S}\rightarrow \mathbf{I})), \label{discrete_s_binom}\\
n(\mathbf{I}\rightarrow \mathbf{R})&\sim \text{Binomial}(\mathbf{I}_{t-1}, p(\mathbf{I}\rightarrow \mathbf{R}))\label{discrete_i_binom},
\end{align}
where $n(\mathbf{S}\rightarrow \mathbf{I})$ and $n(\mathbf{I}\rightarrow \mathbf{R})$ denote the total number of individuals leaving the susceptible and infected compartments, respectively. The complete discrete, stochastic SIR model is therefore presented as
\begin{align}
\mathbf{S}_{t}&=\mathbf{S}_{t-1}-n(\mathbf{S}\rightarrow \mathbf{I}), \label{SIRR:s_discrete}\\
\mathbf{I}_{t}&=\mathbf{I}_{t-1}+ n(\mathbf{S}\rightarrow \mathbf{I}) - n(\mathbf{I}\rightarrow \mathbf{R}),\label{SIRR:i_discrete}\\
\mathbf{R}_{t}&=N_{pop} - \mathbf{S}_{t} - \mathbf{I}_{t}\label{SIRR:r_discrete}.
\end{align}
The likelihood is defined to be 
\begin{align}
\mathbf{y}_{t} \mid \mathbf{x}_{t} \sim \mathcal{P}\left(\mathbf{y}_{t} ;\mathbf{I}_{t}\right),
\end{align}
where $\mathcal{P}$ corresponds to the Poisson distribution. 

The true parameters of the SIR model are $\mathbf{\bm{\theta}} = \{\mathbf{\beta}, \mathbf{\gamma}\} = \{0.85, 0.20\}$ with $T=30$ observations, $N_{pop}=10000$ and $3$ initially infected. The initial samples were drawn from $q_1(\mathbf{\bm{\theta}}^i_1)=U[0,1]$ and propagated according to $q(\mathbf{\bm{\theta}}^i_k|\mathbf{\bm{\theta}}^i_{k-1}) = \mathcal{N}(\mathbf{\bm{\theta}}^i_k; \mathbf{\bm{\theta}}^i_{k-1}, \mathbf{\Sigma})$ where $\mathbf{\Sigma}=0.1\mathbf{I}_2$.  

\subsection{Experimental Setup} \label{sec:setup}
    The analysis conducted in this paper was run on a cluster of $4$ DM machines. Each DM machine consists of a `2 Xeon Gold 6138' CPU, which provides a memory of $384$GB and $40$ cores. This results in a total of $160$ cores. The experiments requested a power-of-two number of cores and treated the cores as DM processors (i.e., $P = 1, 2, 4, \dots, 128$). This is because the parallelization strategy described in Appendix~\ref{app:parallel_resamp} utilizes the divide-and-conquer paradigm. Having a power-of-two number for $P$ optimizes the workload balance across the processors. We also note that with such hardware, one could use a hybrid distributed-shared memory framework. This framework was not implemented here, but in the context of SMC$^2$ could be an interesting direction for future work. 

    For the same reasoning as choosing $P$, $N\in\{256, 512, 1024\}$ is also chosen to be a power-of-two to balance the workload across DM processors. The number of SMC sampler iterations is set to $K=10$ and the number of particles in each PF instance to $N_x = 500$. For consistency of comparison, the number of p-MCMC samples is set to $M = K \times N$ (half of which are burned-in), such that the problem size for both p-MCMC and SMC$^2$ is identical. The results presented in Table \ref{Table:SIR_results} and Figures \ref{fig: runtimes} and \ref{fig: speedups} are averaged over 10 Monte Carlo runs. The source code is available on \url{https://github.com/j-j-murphy/O-logN-SMC-Squared-on-MPI}.
    
\subsection{Discussion} \label{sec:discussion}
\begin{figure}
    \centering
    \includegraphics[width=0.72\linewidth]{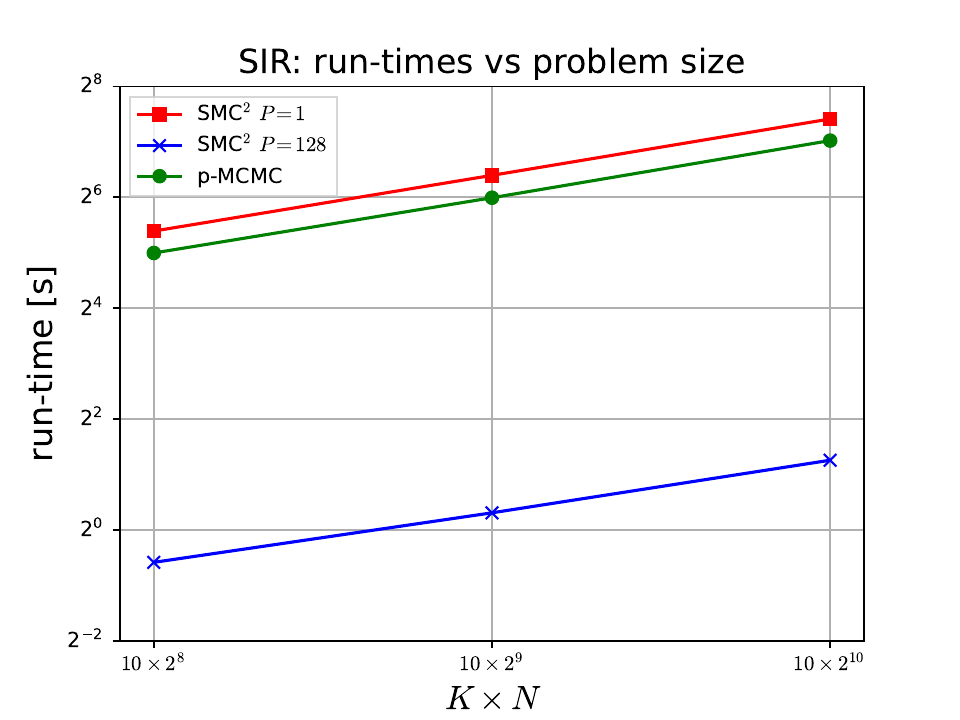}
    \caption{SIR: run-times for increasing problem size, $K \times N$.}
    \label{fig: runtimes}
\end{figure}

\begin{figure}
    \centering
    \includegraphics[width=0.72\linewidth]{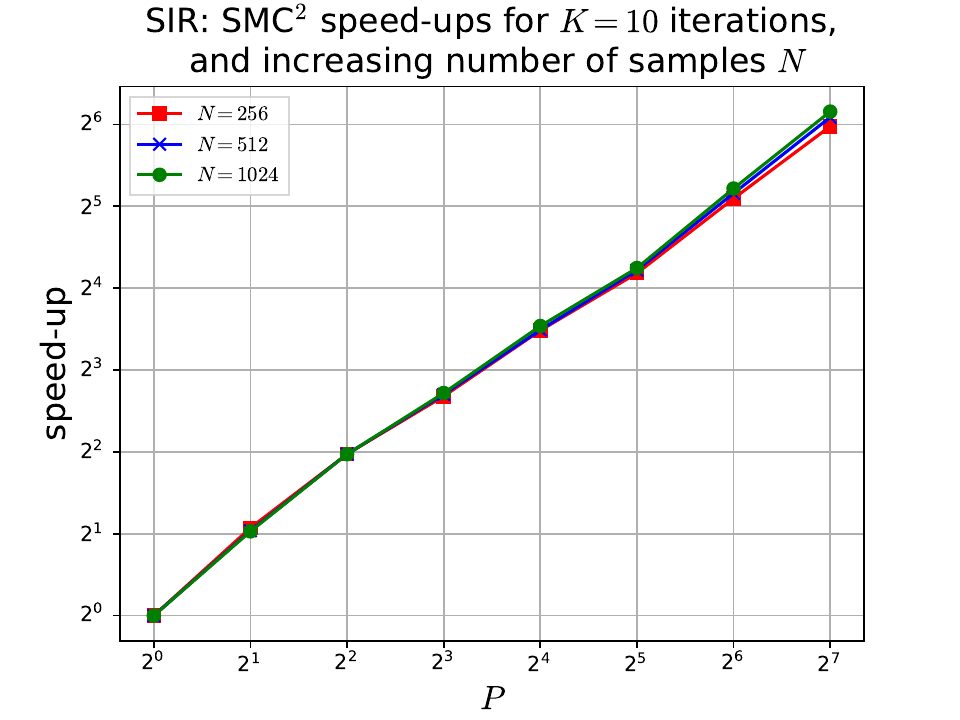}
    \caption{SIR: speed-ups for SMC$^2$ vs DM processors, $P$.}
    \label{fig: speedups}
\end{figure}

\begin{figure}
    \centering
    \begin{subfigure}[t]{\linewidth}
        \centering
        \includegraphics[width=0.78\linewidth]{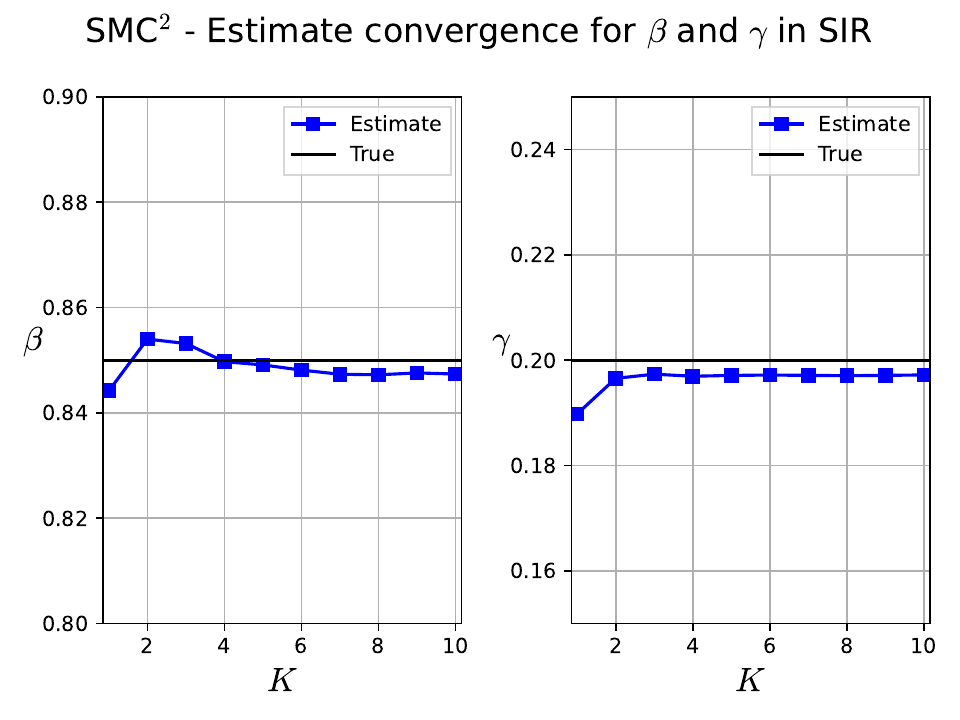}\label{fig:convergence_smc2}
        \caption{SMC$^2$}
    \end{subfigure}
    \begin{subfigure}[t]{\linewidth}
        \centering
        \includegraphics[width=0.78\linewidth]{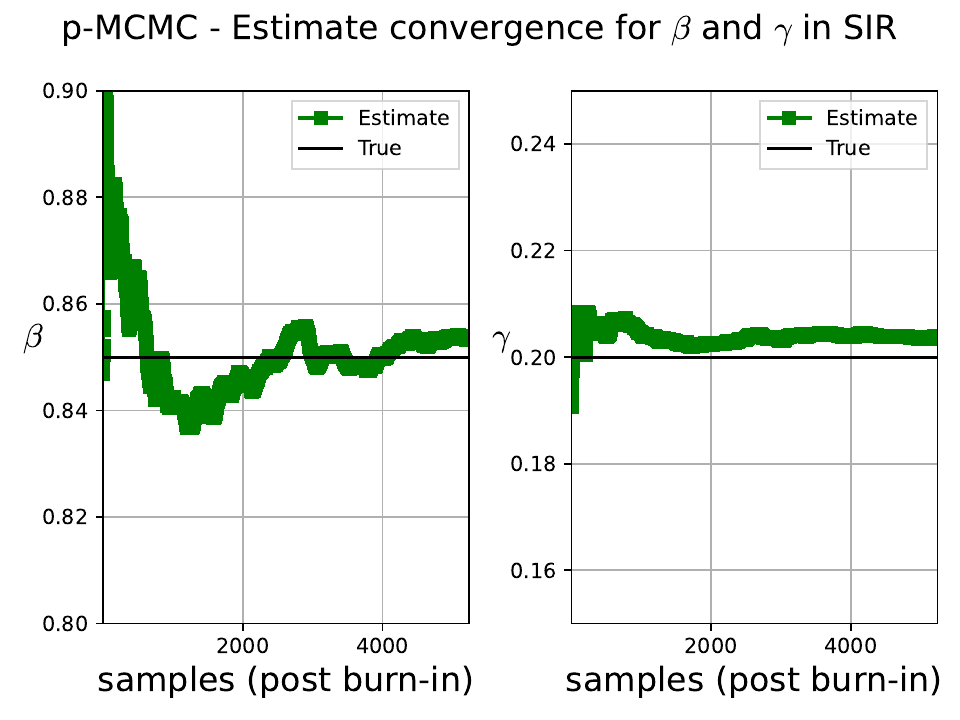}\label{fig:convergence_pmcmc}
        \caption{p-MCMC}
    \end{subfigure}
    \caption{SIR: estimate convergence for $\beta$ and $\gamma$ on a single Monte Carlo run.}
    \label{fig:convergence}
\end{figure}

    The results are available in Figures \ref{fig: runtimes}, \ref{fig: speedups}, \ref{fig:convergence}, and Table \ref{Table:SIR_results}. 

    It is evident that p-MCMC is faster than the sequential version of SMC$^2$ for the same problem size. This is unsurprising as the constant time per sample in p-MCMC only consists of sampling, computing \eqref{eq:MHalgorithm_} and performing the accept/reject step. In contrast, the constant time per sample in SMC$^2$ is given by IS, estimates, and resampling which are more computationally intensive. However, SMC$^2$ is parallelizable while p-MCMC is not. By using $P = 128$ DM processors, SMC$^2$ becomes up to $54\times$ faster than p-MCMC (as shown in Figure \ref{fig: runtimes} and Table \ref{Table:SIR_results}). SMC$^2$ also achieves roughly a $71$-fold speed-up with respect to its sequential version (as illustrated in Figure \ref{fig: speedups}). 

    Table \ref{Table:SIR_results} provides the estimation results for both methods for the highest considered problem size, $K \times N = 10240$. Results for lower values of $K \times N$ are omitted for brevity, due to page limit, and because they are known from theory to achieve lower accuracy. Both methods provide a fairly accurate estimate of both parameters $\beta$ and $\gamma$. However, the proposed SMC$^2$ seems to be statistically more accurate, i.e., with a Mean Squared Error (MSE) roughly three times lower (see Table \ref{Table:SIR_results}), and quicker at converging (as exemplified in Figure \ref{fig:convergence}). We perceive this to be the result of the proposed approach being equipped with efficient and nearly-optimal components, i.e., recycling and the backward kernel, which p-MCMC does not utilize. 

\begin{table}[]
\caption{SIR: Parameter estimation results for the maximum considered problem size, $K \times N = 10240$ samples. The true values for parameters to estimates are: $\beta = 0.85$, $\gamma = 0.2$. }
\renewcommand{\arraystretch}{1.2}
\centering
\begin{tabular}{cccccc}
\hline 
\hline 
\textbf{Method} & $P$ & $E[\beta]$ & $E[\gamma]$ & \textbf{MSE} & \textbf{Time [s]}\\
\hline 
SMC$^2$ & $128$ & $0.8526$ & $0.2003$ & $7.75\times 10^{-5}$ & $2.39$ \\
P-MCMC & - & $0.8628$ & $0.2038$ & $2.42\times10^{-4}$ & $130.01$ \\
\hline \hline
\end{tabular}
\label{Table:SIR_results}
\end{table}

%% file: 7_Conclusions.tex
\section{Conclusions and Future Work}\label{sec:conclusions}
In this paper, we have proposed a high-performance SMC$^2$ algorithm for parameter estimation that is parallelized in optimal time complexity $\mathcal{O}(\log_2N)$. We also include an approximately optimal backward kernel and an efficient recycling scheme. Inferring the parameters of a compartmental disease model on a cluster of $128$ DM processors, SMC$^2$ achieves $\sim71$ speed-up when compared to one processor. SMC$^2$ also provides more accurate estimates than p-MCMC but $54$ times faster. This makes SMC$^2$ an appealing alternative to p-MCMC.

For both SMC$^2$ and p-MCMC we have used RW proposals which make them compatible with both discrete and continuous models. However, better proposals could be utilized. For discrete models, one could use HINTS \cite{HINTS}. Gradient-based proposals that have been used within SMC samplers, such as HMC \cite{smchmc} and NUTS \cite{devlin2021no} could be used for continuous models. The gradient of the log-likelihood w.r.t $\mathbf{\bm{\theta}}$ from a PF would need to be calculated. It has been documented in the literature that the operations inherent to the PF are non-differentiable \cite{rosato2022efficient}. A recent review outlined numerous attempts to overcome these issues \cite{chen2023overview} including the method described in \cite{rosato2022efficient} which uses NUTS as the proposal within p-MCMC. A comparison between p-MCMC and SMC$^2$ with gradient-based proposals would therefore be sensible. We note that these improvements are likely to increase the run-time of SMC$^2$ which could be compensated for by extending the DM parallelization outlined in this paper to hybrid memory e.g., by using clusters of DM GPUs. 

\section*{Acknowledgment}

The authors would like to thank Matthew Carter for his helpful correspondence. 

%% file: 8_References.tex
\bibliographystyle{ieeetr}
\bibliography{bibliography.bib}

%% file: 9_Appendix.tex
\appendices
\balance
\section{$\mathcal{O}(\log_2N)$ Fully-Balanced Parallel SMC$^2$ on DM}	\label{app:parallel_resamp}
\begin{figure}
    \centering
    \includegraphics[width=0.9\linewidth]{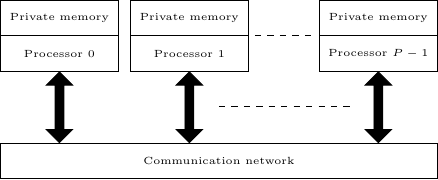}
    \caption{Distributed Memory Architecture}
    \label{fig: DM}
\end{figure}
In this paper, we only consider parallel implementations of SMC$^2$ on DM. DM systems are made of $P$ processors, each being uniquely identified with an integer rank $p = 0, 1, \dots, P-1$. Each processor owns a private memory that contains data that can only be exchanged to other processors' memory through a common communication network exemplified by Figure~\ref{fig: DM}. The analysis undertaken in this paper is coded in Python and parallelized using mpi4py \cite{dalcin2011parallel, dalcin2021mpi4py} which is a commonly used DM framework. Here, we briefly explain how to parallelize each sub-task of SMC$^2$ on DM architecture.

The equations inherent to IS, \eqref{eq: init_sample}, \eqref{init_weights}, \eqref{eq: proposal}, and $\eqref{sec:l_kernels}$ are \textit{embarrassingly parallel} as the iterations are independent $\forall i$. These tasks are trivially parallelized by partitioning the iteration space equally among processors. Each processor, $p$, will allocate space in its private memory and update $n = \frac{N}{P}$ samples and weights with index $i = np+1, np+2, np+3, \dots, n(p+1)$. These tasks achieve $\mathcal{O}(\frac{N}{P}) \rightarrow \mathcal{O}(1)$ time complexity. The \textit{embarrassingly parallel} tasks within resampling include \eqref{eq: ncopies} and the final reset step for the normalized weights.

Equations \eqref{realised_estimates}, \eqref{eq: normalise_smc2}, \eqref{eq: recycling_I}, and \eqref{eq: ess_smc2} require the computation of a vector sum which is parallelizable through parallel reduction operations. These are known to achieve $\mathcal{O}(\frac{N}{P} + \log_2P) \rightarrow \mathcal{O}(\log_2N)$ time complexity. 

Equation \eqref{eq: cdf} can be parallelized by using the parallel prefix sum operation, also known as cumulative sum or scan. The achieved time complexity is also $\mathcal{O}(\frac{N}{P} + \log_2P) \rightarrow \mathcal{O}(\log_2N)$ which was proven to be optimal in~\cite{Cumulative_Sum}.

Since \eqref{eq: ncopies_definition} holds, Algorithm \ref{alg: sequential_redistribution} also takes $\mathcal{O}(N)$ iterations for $P = 1$. However, it is not possible to parallelize this task in an \textit{embarrassingly parallel} fashion due to the workload being randomly distributed across the DM processors. As \eqref{eq: ncopies_range} explains, the workload may be strongly unbalanced. In \cite{Alessandro5}, it is shown that it is possible to parallelize redistribution on DM in optimal $\mathcal{O}(\log_2N)$ time complexity. This approach splits redistribution into three steps: Rotational Nearly Sort, Rotational Split, and S-R (see Algorithm \ref{alg: sequential_redistribution}). This Appendix briefly summarizes the main points of each step. 

\textit{Rotational Nearly Sort}. The aim of this step is to move $\mathbf{\bm{\theta}}$, and the corresponding elements in $\mathbf{ncopies}$ to lower memory locations (i.e., lower values of the index $i$) until $\mathbf{ncopies}$ has the following shape:
\begin{equation} \label{eq: nearly_sort}
    \mathbf{ncopies} = \left[\mathbf{\bm{\lambda}}^1, \mathbf{\bm{\lambda}}^2, \dots, \mathbf{\bm{\lambda}}^{m}, 0, 0, \dots, 0\right],
\end{equation}  
where $\mathbf{\bm{\lambda}}^i$ $\forall i$ are positive integers.
This is done by using prefix sum to compute $\mathbf{shifts} \in \mathbb{Z}^N$ which is an array whose $i$-th element, $\mathbf{shift}^i$, counts how many $\mathbf{ncopies}^i = 0$ are found up to $i$-th position. The values in $\mathbf{shift}$ are expressed as base-$2$ numbers. All $\mathbf{\bm{\theta}}^i$ that have $\mathbf{ncopies}^i > 0$ are rotated up to $\log_2P$ times to lower memory locations. This may result in them being set to lower ranked processors. The number of positions increases by a power-of-$2$, starting from $1$, if and only if the Least Significant Bit (LSB) of $\mathbf{shifts}^i$ is $1$.

As previously mentioned, the prefix sum to compute $\mathbf{shifts}$ can be performed in parallel in $\mathcal{O}(\log_2N)$ time. Each processor sends and/or receives up to $\frac{N}{P}$ samples (and elements of $\mathbf{ncopies}$ and $\mathbf{shifts}$) up to $\log_2P$ times. Therefore, this step achieves $\mathcal{O}(\frac{N}{P}\log_2P) \rightarrow \mathcal{O}(\log_2N)$ time complexity. 

\textit{Rotational Split}. In contrast to Rotational Nearly Sort, this step moves $\mathbf{\bm{\theta}}$ and its respective elements in $\mathbf{ncopies}$ to higher memory locations (i.e., higher values of the index, $i$) until the total workload (represented by the samples to be redistributed) is fully balanced across the DM processor. This occurs when:
\begin{subequations}
    \vspace{-3ex}
    \begin{align} \label{eq: ncopies_after_split}
        \mathbf{ncopies} = [\mathbf{\bm{\lambda}}^1, 0,\dots,0, \mathbf{\bm{\lambda}}^2, 0,\dots,0, \mathbf{\bm{\lambda}}^{m}, 0,\dots,0] \\
    \sum\nolimits_{i = p \frac{N}{P}+1}^{(p+1) \frac{N}{P}} \mathbf{ncopies}^i = \frac{N}{P}  \ \ \  \forall p = 0, \dots, P-1 \label{eq: workload balanced},
    \end{align}
\end{subequations}
where for each $\mathbf{ncopies}^i > 0$ ($\mathbf{\bm{\lambda}}$s in \eqref{eq: ncopies_after_split}), $\mathbf{ncopies}^i - 1$ zeros follow. Initially, Rotational Split computes $\mathbf{csum} \in \mathbb{Z}^N$ which is the cumulative sum of $\mathbf{ncopies}$. Each element in $\mathbf{csum}$, i.e., $\mathbf{csum}^i$, can be used to calculate how many positions each copy of $\mathbf{\bm{\theta}}^i$ has to rotate to achieve \eqref{eq: ncopies_after_split}. It is possible to prove that the copies of $\mathbf{\bm{\theta}}^i$ will have to move by a minimum of
\begin{equation} \label{eq: min_shifts}
    \mathbf{min\textunderscore shifts}^i = 
    \mathbf{csum}^i - \mathbf{ncopies}^i - i,
\end{equation}
and maximum of
\begin{equation} \label{eq: max_shifts}
    \mathbf{max\textunderscore shifts}^i = \mathbf{csum}^i-i-1.
\end{equation} 
As is with Rotational Nearly Sort, the elements in $\mathbf{min\textunderscore shifts}$ and $\mathbf{max\textunderscore shifts}$ will be expressed in base$-2$. Each $\mathbf{\bm{\theta}}^i$, and its corresponding number of copies, $\mathbf{ncopies}^i$, are rotated in parallel by up to $\log_2P$ times to higher memory locations (and potentially sent to a higher ranked processor). Each iteration decreases the number of positions by power-of-$2$ starting from $\frac{N}{2}$. This is dependent on the Most Significant Bit (MSBs) of $\mathbf{min\textunderscore shifts}^i$ and $\mathbf{max\textunderscore shifts}^i$. 

For obvious reasons, $\mathbf{\bm{\theta}}^i$ and $\mathbf{ncopies}^i$ must not rotate if the MSBs of $\mathbf{min\textunderscore shifts}^i$ and $\mathbf{max\textunderscore shifts}^i$ are $0$. Trivially, if the MSBs of $\mathbf{min\textunderscore shifts}^i$ and $\mathbf{max\textunderscore shifts}^i$ are $1$, all copies of $\mathbf{\bm{\theta}}^i$ must rotate. However, if the MSB of $\mathbf{max\textunderscore shifts}^i$ is $1$, some copies must not move and the remaining must split and rotate. Those for which $\mathbf{csum}^i-i$ is higher than the power-of-$2$ number of positions to rotate by at any given iteration. After $\log_2P$ iterations, \eqref{eq: workload balanced} will hold, meaning each processor will have $\frac{N}{P}$ samples to redistribute.

Computing $\mathbf{csum}$ for the first time involves a parallel prefix sum which incurs a cost of $\mathcal{O}(\log_2N)$. However, if the elements in $\mathbf{\bm{\theta}}$ and $\mathbf{ncopies}$ are moved, the elements of $\mathbf{csum}$ move consequently. This avoids recomputing $\mathbf{csum}$ from scratch, which would cost $\mathcal{O}(\log_2N)$. Therefore, $\mathbf{csum}$ can be updated in $\mathcal{O}(\frac{N}{P}) \rightarrow \mathcal{O}(1)$. Similar to Rotational Nearly Sort, each processor sends and/or receives up to $\frac{N}{P}$ samples (and elements of $\mathbf{ncopies}$ and $\mathbf{csum}$) up to $\log_2P$ times. Therefore, this step also achieves $\mathcal{O}(\frac{N}{P}\log_2P) \rightarrow \mathcal{O}(\log_2N)$ time complexity.

\textit{S-R}. After performing Rotational Split, all processors have the same number of samples to redistribute. Therefore, each processor can call Algorithm \ref{alg: sequential_redistribution} over its subset of $\frac{N}{P}$ samples. This is completed in $\mathcal{O}(\frac{N}{P}) \rightarrow \mathcal{O}(1)$ iterations.
\begin{figure}
    \centering
    \includegraphics[width=0.8\linewidth]{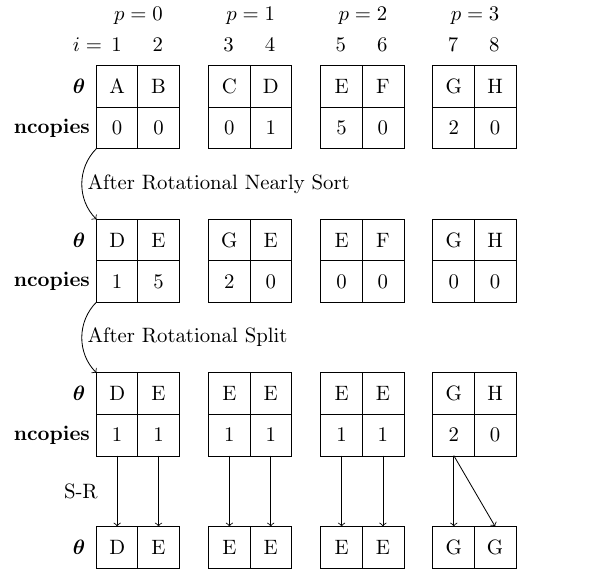}
    \caption{Redistribution - Example for $N = 8$ and $P = 4$. The values of $\mathbf{\bm{\theta}}$ are typically vectors of numbers, but here they are represented as letters for brevity.}
    \label{fig: redistribution}
\end{figure}
\begin{algorithm}
    \small
    \caption{Parallel Redistribution}\label{alg: parallel_redistribution}
    \textbf{Input: } $\mathbf{\bm{\theta}}$, $\mathbf{ncopies}$, $N$, $P$ \\
    \textbf{Output: } $\mathbf{\bm{\theta}}$
    \begin{algorithmic}[1]
	\If{$P > 1$}
            \State $\mathbf{\bm{\theta}}, \mathbf{ncopies} \gets$ \texttt{Rot. Nearly Sort}$(\mathbf{\bm{\theta}}, \mathbf{ncopies}, N, P)$
            \State $\mathbf{\bm{\theta}}, \mathbf{ncopies} \gets$ \texttt{Rotational Split}$(\mathbf{\bm{\theta}}, \mathbf{ncopies}, N, P)$
        \EndIf 
        \State $\mathbf{\bm{\theta}} \gets$\texttt{S-R}$(\mathbf{\bm{\theta}}, \mathbf{ncopies}, N)$
    \end{algorithmic}
\end{algorithm}
Figure \ref{fig: redistribution} illustrates a practical example with $N=8$ samples and $P=4$ processors. Algorithm \ref{alg: parallel_redistribution} summarizes the steps for parallel redistribution. The following theorem and remark prove that the described parallelization approach for SMC$^2$ has asymptotically optimal for $P \rightarrow N$ time complexity. 

\begin{thm} \label{thm: optimality}
    Let $\mathbf{\bm{\theta}}_k$ be an array of $N$ (multi-dimensional) samples that need to be resampled according to the normalized weights in $\mathbf{\Tilde{w}}_k \in \mathbb{R}^N$. Let the elements of $\mathbf{\bm{\theta}}_k$ and $\mathbf{\Tilde{w}}_k$ be equally distributed on a DM architecture across $P$ DM processors. A resampling algorithm performing \eqref{eq: cdf}, \eqref{eq: ncopies}, Algorithm \ref{alg: parallel_redistribution} and the reset of all values in $\mathbf{\Tilde{w}}_k$ to $\frac{1}{N}$ in parallel achieves asymptotically optimal time complexity equal to $\mathcal{O}(\log_2N)$.
\end{thm}
\begin{proof}
    Overall, parallel redistribution in Algorithm \ref{alg: parallel_redistribution} achieves $\mathcal{O}(\log_2N)$ time complexity. This is calculated by summing the time complexity of each process (i.e., Rotational Nearly Sort, Rotational Split, and S-R) and taking the limit for $P \rightarrow N$. It is proven in \cite{Cumulative_Sum}, a parallel implementation of \eqref{eq: cdf} is optimally bound to $\mathcal{O}(\log_2N)$. Equation \eqref{eq: ncopies} and the resetting of the normalized weights costs $\mathcal{O}(\frac{N}{P}) \rightarrow \mathcal{O}(1)$. 
    
     If it were possible to perform Theorem \ref{thm: optimality} in less than $\mathcal{O}(\log_2N)$, (e.g., in $\mathcal{O}(1)$), the overall time complexity of resampling would still be bound to $\mathcal{O}(\log_2N)$ by \eqref{eq: cdf}. 
\end{proof}
\begin{rem}
    We note that Theorem \ref{thm: optimality} means that SMC$^2$ is asymptotically optimal as the overall time complexity of each iteration of SMC$^2$ is bound to the one of resampling. This is summarized in Table~\ref{tab: smc2_tasks}.
\end{rem}

%% file: conference_101719.bbl
\begin{thebibliography}{10}

\bibitem{del2006sequential}
P.~Del~Moral, A.~Doucet, and A.~Jasra, ``Sequential monte carlo samplers,''
  {\em Journal of the Royal Statistical Society: Series B (Statistical
  Methodology)}, vol.~68, no.~3, pp.~411--436, 2006.

\bibitem{altekar2004parallel}
G.~Altekar, S.~Dwarkadas, J.~P. Huelsenbeck, and F.~Ronquist, ``Parallel
  metropolis coupled markov chain monte carlo for bayesian phylogenetic
  inference,'' {\em Bioinformatics}, vol.~20, no.~3, pp.~407--415, 2004.

\bibitem{green2022increasing}
P.~L. Green, L.~Devlin, R.~E. Moore, R.~J. Jackson, J.~Li, and S.~Maskell,
  ``Increasing the efficiency of sequential monte carlo samplers through the
  use of approximately optimal l-kernels,'' {\em Mechanical Systems and Signal
  Processing}, vol.~162, p.~108028, 2022.

\bibitem{hastings1970monte}
W.~K. Hastings, ``Monte carlo sampling methods using markov chains and their
  applications,'' 1970.

\bibitem{robert2010metropolis}
C.~Robert, G.~Casella, C.~P. Robert, and G.~Casella, ``Metropolis--hastings
  algorithms,'' {\em Introducing Monte Carlo Methods with R}, pp.~167--197,
  2010.

\bibitem{neal2012mcmc}
R.~M. Neal {\em et~al.}, ``Mcmc using hamiltonian dynamics,'' {\em Handbook of
  markov chain monte carlo}, vol.~2, no.~11, p.~2, 2011.

\bibitem{NUTS}
M.~D. Hoffman, A.~Gelman, {\em et~al.}, ``The no-u-turn sampler: adaptively
  setting path lengths in hamiltonian monte carlo.,'' {\em J. Mach. Learn.
  Res.}, vol.~15, no.~1, pp.~1593--1623, 2014.

\bibitem{pymc3}
J.~Salvatier, T.~V. Wiecki, and C.~Fonnesbeck, ``Probabilistic programming in
  python using pymc3,'' {\em PeerJ Computer Science}, vol.~2, p.~e55, 2016.

\bibitem{carpenter2017stan}
B.~Carpenter, A.~Gelman, M.~D. Hoffman, D.~Lee, B.~Goodrich, M.~Betancourt,
  M.~Brubaker, J.~Guo, P.~Li, and A.~Riddell, ``Stan: A probabilistic
  programming language,'' {\em Journal of statistical software}, vol.~76,
  no.~1, pp.~1--32, 2017.

\bibitem{arulampalam2002tutorial}
M.~S. Arulampalam, S.~Maskell, N.~Gordon, and T.~Clapp, ``A tutorial on
  particle filters for online nonlinear/non-gaussian bayesian tracking,'' {\em
  IEEE Transactions on signal processing}, vol.~50, no.~2, pp.~174--188, 2002.

\bibitem{resamplingmethods}
J.~D. Hol, T.~B. Schon, and F.~Gustafsson, ``On resampling algorithms for
  particle filters,'' in {\em 2006 IEEE nonlinear statistical signal processing
  workshop}, pp.~79--82, IEEE, 2006.

\bibitem{Lopez}
F.~Lopez, L.~Zhang, J.~Beaman, and A.~Mok, ``Implementation of a particle
  filter on a gpu for nonlinear estimation in a manufacturing remelting
  process,'' in {\em 2014 IEEE/ASME International Conference on Advanced
  Intelligent Mechatronics}, pp.~340--345, July 2014.

\bibitem{Lopez2}
F.~Lopez, L.~Zhang, A.~Mok, and J.~Beaman, ``Particle filtering on gpu
  architectures for manufacturing applications,'' {\em Computers in Industry},
  vol.~71, pp.~116 -- 127, 2015.

\bibitem{Murray_Multiresampling}
L.~M. Murray, A.~Lee, and P.~E. Jacob, ``Parallel resampling in the particle
  filter,'' {\em Journal of Computational and Graphical Statistics}, vol.~25,
  no.~3, pp.~789--805, 2016.

\bibitem{Alessandro2}
A.~Varsi, J.~Taylor, L.~Kekempanos, E.~Pyzer~Knapp, and S.~Maskell, ``A fast
  parallel particle filter for shared memory systems,'' {\em IEEE Signal
  Processing Letters}, vol.~27, pp.~1570--1574, 2020.

\bibitem{Alessandro}
A.~Varsi, L.~Kekempanos, J.~Thiyagalingam, and S.~Maskell, ``Parallelising
  particle filters with deterministic runtime on distributed memory systems,''
  {\em IET Conference Proceedings}, pp.~11--18, 2017.

\bibitem{Alessandro3}
A.~Varsi, L.~Kekempanos, J.~Thiyagalingam, and S.~Maskell, ``A single smc
  sampler on mpi that outperforms a single mcmc sampler",'' {\em eprint
  arXiv:1905.10252}, 2019.

\bibitem{Alessandro5}
A.~Varsi, S.~Maskell, and P.~G. Spirakis, ``An o (log2n) fully-balanced
  resampling algorithm for particle filters on distributed memory
  architectures,'' {\em Algorithms}, vol.~14, no.~12, pp.~342--362, 2021.

\bibitem{andrieu_doucet_holenstein_2010}
C.~Andrieu, A.~Doucet, and R.~Holenstein, ``Particle markov chain monte carlo
  methods,'' {\em Journal of the Royal Statistical Society: Series B
  (Statistical Methodology)}, vol.~72, no.~3, pp.~269--342, 2010.

\bibitem{chopin2013smc2}
N.~Chopin, P.~E. Jacob, and O.~Papaspiliopoulos, ``Smc2: an efficient algorithm
  for sequential analysis of state space models,'' {\em Journal of the Royal
  Statistical Society: Series B (Statistical Methodology)}, vol.~75, no.~3,
  pp.~397--426, 2013.

\bibitem{carson2018bayesian}
J.~Carson, M.~Crucifix, S.~Preston, and R.~D. Wilkinson, ``Bayesian model
  selection for the glacial--interglacial cycle,'' {\em Journal of the Royal
  Statistical Society Series C: Applied Statistics}, vol.~67, no.~1,
  pp.~25--54, 2018.

\bibitem{ABM}
L.~Rimella, S.~Alderton, M.~Sammarro, B.~Rowlingson, D.~Cocker, N.~Feasey,
  P.~Fearnhead, and C.~Jewell, ``{Inference on extended-spectrum beta-lactamase
  Escherichia coli and Klebsiella pneumoniae data through SMC2},'' {\em Journal
  of the Royal Statistical Society Series C: Applied Statistics}, p.~qlad055,
  07 2023.

\bibitem{lux2023sequential}
T.~Lux, ``Sequential monte carlo squared for agent-based models,'' in {\em
  Artificial Intelligence, Learning and Computation in Economics and Finance},
  pp.~59--69, Springer, 2023.

\bibitem{nguyen2015efficient}
T.~L.~T. Nguyen, F.~Septier, G.~W. Peters, and Y.~Delignon, ``Efficient
  sequential monte-carlo samplers for bayesian inference,'' {\em IEEE
  Transactions on Signal Processing}, vol.~64, no.~5, pp.~1305--1319, 2015.

\bibitem{doucet2001sequential}
A.~Doucet, N.~De~Freitas, N.~J. Gordon, {\em et~al.}, {\em Sequential Monte
  Carlo methods in practice}, vol.~1.
\newblock Springer, 2001.

\bibitem{rosato2022efficient}
C.~Rosato, L.~Devlin, V.~Beraud, P.~Horridge, T.~B. Sch{\"o}n, and S.~Maskell,
  ``Efficient learning of the parameters of non-linear models using
  differentiable resampling in particle filters,'' {\em IEEE Transactions on
  Signal Processing}, vol.~70, pp.~3676--3692, 2022.

\bibitem{kermack1927contribution}
W.~O. Kermack and A.~G. McKendrick, ``A contribution to the mathematical theory
  of epidemics,'' {\em Proceedings of the royal society of london. Series A,
  Containing papers of a mathematical and physical character}, vol.~115,
  no.~772, pp.~700--721, 1927.

\bibitem{HINTS}
M.~Strens, ``Efficient hierarchical mcmc for policy search,'' in {\em
  Proceedings of the Twenty-First International Conference on Machine
  Learning}, ICML '04, (New York, NY, USA), pp.~97--104, Association for
  Computing Machinery, 2004.

\bibitem{smchmc}
A.~Buchholz, N.~Chopin, and P.~Jacob, ``Adapative tuning of hamiltonian monte
  carlo within sequential monte carlo,'' {\em Bayesian Analysis}, vol.~16,
  pp.~745--771, 2021.

\bibitem{devlin2021no}
L.~Devlin, P.~Horridge, P.~L. Green, and S.~Maskell, ``The no-u-turn sampler as
  a proposal distribution in a sequential monte carlo sampler with a
  near-optimal l-kernel,'' {\em arXiv preprint arXiv:2108.02498}, 2021.

\bibitem{chen2023overview}
X.~Chen and Y.~Li, ``An overview of differentiable particle filters for
  data-adaptive sequential bayesian inference,'' {\em arXiv preprint
  arXiv:2302.09639}, 2023.

\bibitem{dalcin2011parallel}
L.~D. Dalcin, R.~R. Paz, P.~A. Kler, and A.~Cosimo, ``Parallel distributed
  computing using python,'' {\em Advances in Water Resources}, vol.~34, no.~9,
  pp.~1124--1139, 2011.

\bibitem{dalcin2021mpi4py}
L.~Dalcin and Y.-L.~L. Fang, ``mpi4py: Status update after 12 years of
  development,'' {\em Computing in Science \& Engineering}, vol.~23, no.~4,
  pp.~47--54, 2021.

\bibitem{Cumulative_Sum}
E.~E. Santos, ``Optimal and efficient algorithms for summing and prefix summing
  on parallel machines,'' {\em Journal of Parallel and Distributed Computing},
  vol.~62, no.~4, pp.~517--543, 2002.

\end{thebibliography}
